\documentclass{article}

\usepackage{arxiv}

\usepackage{amsfonts}       %
\usepackage{amsmath,amssymb}
\usepackage{booktabs}       %
\usepackage{color}
\usepackage{enumitem}
\usepackage{epsfig}
\usepackage{subcaption}
\usepackage{graphicx}
\usepackage{lineno}
\usepackage[linesnumbered,ruled,vlined]{algorithm2e}
\usepackage{lipsum}
\usepackage{mathrsfs}
\usepackage{microtype}      %
\usepackage{multirow}
\usepackage{nicefrac}       %
\usepackage{rotating}
\usepackage[T1]{fontenc}    %
\usepackage{url}            %
\usepackage[utf8x]{inputenc} %

\usepackage{listings}
\usepackage{pmboxdraw}
\usepackage{url}
\usepackage{hyperref}
\usepackage{tcolorbox}
\usepackage{tikz}
\usepackage{amsmath}
\usepackage{pgfplots}
\pgfplotsset{width=10cm,compat=1.17}
\usetikzlibrary{calc}

  \title{Mathematical Content Browsing for Print-Disabled Readers Based on Virtual-World Exploration and Audio-Visual Sensory Substitution}

\author{
  Rynhardt Kruger, Febe de Wet, Thomas ~Niesler \\
  Department of Electrical and Electronic Engineering\\
  University of Stellenbosch\\
  Stellenbosch,  South Africa \\
  \texttt{rkruger@sun.ac.za}, \texttt{fdw@sun.ac.za}, \texttt{trn@sun.ac.za} \\
}

\setcitestyle{square,numbers}
\begin{document}

\maketitle
\begin{abstract} %
Documents containing mathematical content remain largely inaccessible to blind and visually impaired readers because they are predominantly published as untagged PDF which does not include the semantic data necessary for effective accessibility.
Equations in such documents consist of images interlaced with text, and cannot be interpreted using a screen reader.
We present a browsing approach for print-disabled readers specifically aimed at such mathematical content. 
This approach draws on the navigational mechanisms often used to explore the virtual worlds of text adventure games with audio-visual sensory substitution for graphical content.
The relative spatial placement of the elements of an equation are represented as a virtual world, so that the reader can navigate from element to element.
Text elements are announced conventionally using synthesised speech while graphical elements, such as roots and fraction lines, are rendered using a modification of the vOICe algorithm.
The virtual world allows the reader to interactively discover the spatial structure of the equation, while the rendition of graphical elements as sound allows the shape and identity of elements that cannot be synthesised as speech to be discovered and recognised.
The browsing approach was evaluated by eleven blind and fourteen sighted participants in a user trial that included the identification of twelve equations extracted from PDF documents.
Overall, equations were identified completely correctly in 78\% of cases (74\% and 83\% respectively for blind and sighted subjects).
If partial correctness is considered, the performance is substantially higher.
Feedback from the blind subjects indicated that the technique allows spatial information and graphical detail to be discovered
that was hitherto unavailable to them, and that this was useful when exploring the equations.
We conclude that the integration of a spatial model represented as a virtual world in conjunction with audio-visual sensory substitution for non-textual elements can be an effective way for blind and visually impaired readers to read currently inaccessible mathematical content in PDF documents.
  \end{abstract}
\section{Introduction} 
The proliferation of electronic text has made it possible for blind and print-disabled readers to access articles and other documents directly without first requiring conversion from physically printed material. 
This is achieved by the use of screen readers and other assistive technology, which can convey electronic text to a blind reader via alternative mediums, such as synthesised speech and electronic braille. 
However, such mediums are currently unable to convey graphical information like diagrams and print mathematics, which are by nature two-dimensional.
Although standards exist for encoding accessible representations of mathematical equations, diagrams, and other graphical material, in practice most published work does not adhere to these standards. 
In particular, most scientific and technical papers are published as electronic documents in the portable document format (PDF) without additional accessibility information. 
For these documents, screen readers provide access only to the textual content. 
Equations and diagrams, which are often key to a proper understanding of the work, remain inaccessible.

We present a method that makes typeset equations, as typically found in scientific papers published as PDFs, directly accessible to blind and visually impaired readers by means of interactive textual exploration and audio sensory substitution. 
Such equations are a mixture of alphanumeric and graphical information, and we consider how they can be explored by a blind reader to discover their constituent parts, both textual and graphical, as well as the spatial relationships between these elements.
Our work builds on the vOICe algorithm proposed by Meijer~\cite{meijer1992experimental}, which encodes visual information as sound, and combines this with approaches used in text-based adventure games to interactively navigate spatially-structured information. 
To some extent, the current study builds on our own previous work in which we proposed extensions to the vOICe that allows it to be used for the interactive exploration of simple diagrams using gestures and a touch screen.

In the following section, we review the current state of the art in terms of accessibility of mathematical material to blind readers. 
We then describe our proposed approach, and evaluate it by means of a set of user trials. 
We further discuss our key findings, as well as informal user feedback gathered from the participants. 
We conclude by describing future directions of our research.

\section{Background}
From an accessibility perspective, technical material in digital format can be broadly categorised into two types of content. 
The first is textual content like the text of an article, which can already be accessed by blind and visually-impaired readers using screen readers and other assistive technology~\cite{Evans2008,king2012screenreaders}.
The second is two-dimensional content with graphical elements, such as equations and diagrams.
This type of content can be made accessible via a screen reader if it is encoded using a semantically rich structure, such as MathML for mathematical equations~\cite{mathml}.
However, especially in most currently available PDF documents, such two-dimensional content is most often encoded as vector graphics or as rasterised image data.
In this case, the content is not accessible using a screen reader unless additional accessibility annotations have been included in the document, such as alternative text (``alt text'') as specified by the Web Content Accessibility Guidelines~\cite{wcag2} or the PDF/UA specification~\cite{pdfaccessibility, ISO14289-1, drummer2012pdf}.
The PDF/UA ISO standard prescribes tags that can be added to PDF documents to improve accessibility. 
These tags are similar to HTML markup, and can be used to convey semantic information. 
However, most current technical and scientific papers continue to be published as untagged PDF documents.
It should also be borne in mind that textual and two-dimensional content are often interlaced in a document, for example when the symbols of an equation or the labels of a diagram are represented as electronic text. 
Without the surrounding graphical content to provide context, however, this textual information can generally not be interpreted by a blind or visually impaired reader.

Technical material can be made accessible by presenting the content in an alternative (usually linear) format which can be read using braille or synthesised speech.
This approach is appropriate when the graphical content is a visual representation of semantic information that can be represented in an alternative textual form that is accessible via braille or TTS, as is the case for example for mathematical equations and chemical formulas~\cite{karshmer2002access}. 
Alternatively, accessibility can be achieved by direct translation of the content into a form that can be interpreted by a human sense other than sight.
This approach, known as sensory substitution, is appropriate when the shape of the graphical content itself is important, such as for example in the case of plans and geographical maps~\cite{king2007re}. 
An example of a form of sensory substitution is the use of tactile diagrams, in which the sense of touch is used to convey information that is usually encoded visually~\cite{miller2010guidelines}.

\subsection{Linear representations of graphical content}
Perhaps the most easily understood linear representation for graphical content is a textual description. 
Current accessibility standards, for example, require that graphical information is described in this way to print-disabled readers~\cite{wcag2}. 
This is usually accomplished by adding a special attribute containing the textual description to the information source, a mechanism that is known as alternative text or simply ``alt text''. 
However, the linear nature of textual descriptions may not be appropriate for the description of complex content that includes structured relationships between its constituent parts. 
Furthermore, an adequate textual description for complex graphical information might be impractically long~\cite{larkin1987diagram}.

Several specialised linear formats have been developed for reading mathematical content. 
The best known example is arguably braille mathematics. 
Although different standards exist for braille mathematics, for example Nemeth braille~\cite{nemeth} and UK mathematics~\cite{ukmath}, they all make use of additional markup to provide context that is not readily apparent in a linear form. 
For instance, mathematical braille codes provide markup that indicates the start and end of a fraction.
These demarcations are immediately apparent to a sighted reader looking at a graphical representation of the equation, but much harder to determine from a linear textual representation. 

Similar markup is utilised by linear textual mathematical formats like LaTeX and ASCII Math. 
Although these formats can be read by blind and visually impaired readers, they were designed primarily with typesetting in mind. 
Their braille representations are therefore not as easy to read as braille mathematics codes, and their direct rendition by a screen reader as synthesised speech is cumbersome~\cite{stoger2015accessing,melfi2018inclusive}.
For this reason, a number of ways to provide a more efficient means of reading linear mathematical content with a screen reader have been proposed.

One of the first examples was AsTeR, described by Raman~\cite{raman1994aster}.
AsTeR converts mathematical content in LaTeX format into synthesised speech, making use of different speech attributes to indicate aspects of the context.
For instance, AsTeR will read a numerator in a higher pitch, while a denominator will be read in a lower pitch. 
AsTeR also provides commands for moving forward and backward through the content, as well as exploring structures in more detail. 
With the development of the MathML standard for displaying mathematics on the web, similar speech browsers have been devised for the exploration of MathML using a screen reader. 

The company Design Science Inc develops the Math Player extension for some web browsers. 
This software provides a hierarchical explorer for mathematical equations published in MathML format for screen reader users~\cite{soiffer2005mathplayer,soiffer2007mathplayer2}. 
Equations are represented as a tree structure, with sub-expressions represented as child nodes of their enclosing expressions. 
Similar software extensions are provided for popular screen readers, such as the Access8 extension for the NVDA screen reader~\cite{access8}.
An accessibility mode which includes a hierarchical explorer, was also recently added to the MathJax library used to render TeX equations on the web~\cite{cervone2016employing, cervone2016towards}.
However, all these approaches require the semantic information of the equation to be specified as markup, such as LaTeX or MathML.
In practice, this markup is very rarely present in published electronic documents.

Studying mathematics in a linear way also presents several challenges to a blind or visually impaired person~\cite{karshmer2002access,jayant2006survey}.
Linear formats require the introduction of additional markup to indicate context that is not readily apparent without the spatial information included in a typeset equation.
The result is an increase in the total number of symbols describing the equation. 
Furthermore, since all these symbols are part of one linear sequence, blind and visually impaired readers must determine the context of a term by reading ahead or backwards through the linear representation to find the relevant contextual markup. 
This problem is partially addressed by the hierarchical structure representations described above~\cite{stoger2015accessing}.

The proliferation of several different linear representations for accessing mathematics has also added to the difficulty of communication between blind and sighted readers, as well as between blind readers using different linear representations. 
For instance, several standards exist for the representation of mathematics in braille, but these are not compatible~\cite{stoger2015accessing}. 
Translators have been developed to allow conversion between print mathematics and several linear formats~\cite{karshmer2003uma,thompson2005latex2tri,edwards2006lambda,gardner2014lean}.
However, these translators require access to the semantic information as captured by MathML, for example, and this is often not available.

Hence, although many approaches to the accessibility of mathematical material have been proposed, the mathematical content of almost all published material is represented only as graphical layout, and the semantic information required for accessibility is not available.
The conversion from graphical formats, such as those in untagged PDF documents, to accessible formats, such as braille or MathML, is usually a manual process.
Pattern recognition approaches have been proposed to automate this conversion~\cite{suzuki2003infty,yamaguchi2008new}.  
However, pattern-based approaches can only recognise elements that were included in the training data. 
Furthermore, the output may contain recognition errors, which a blind reader has no means to detect without access to the visual representation~\cite{stoger2015accessing}.

\subsection{Sensory substitution}
An alternative and more direct approach to the provision of access to two-dimensional visual material is by direct translation of the visual data to a form that can be perceived by a different human sense.
This approach is sometimes referred to as sensory substitution, as it substitutes one sense with another for the perception of information. 
Two main sensory substitution methods that have been studied for non-visual accessibility are tactile perception and auditory feedback.

\subsubsection{Tactile perception}
Perhaps the best-known example of tactile-visual sensory substitution is the tactile diagram. 
A tactile diagram can be produced by hand using cardboard and other materials, but can also be produced by braille embossers equipped for this task~\cite{viewplus}, or printed as ink on material which expands when heat is subsequently applied~\cite{swelpaper}. 
Tactile diagrams will often require manual intervention in the conversion process, as the resolution of the human tactile sense is lower than that of the visual sense~\cite{loomis2018sensory}.
Tactile diagrams also require labels and other textual elements to be represented in braille, which takes up more space than the printed equivalent~\cite{miller2010guidelines}. 
Finally, perceiving information via touch necessitates different strategies for accessing and localising information, and may be more vulnerable to systematic distortion~\cite{hatwell2003touching}.

When digital diagrams do include semantic metadata, it is possible to produce tactile diagrams with less manual intervention.
This is accomplished by offering textual descriptions of graphical components alongside the tactile diagram, based on the metadata contained in the digital diagram. 
The company View Plus offers a product called Iveo, which enables a blind reader to interactively study an embossed diagram by placing it on a touch pad~\cite{iveo}. 
Although the lower resolution of the human tactile sense limits the effectiveness of this approach, it does allow an interactive spatial exploration of the diagram.

A disadvantage of tactile technology is its significant cost to the user. 
Tactile diagram production requires specialised hardware, and often manual editing, both of which are expensive.
Alternative and lower-cost hardware has been proposed for tactile representation.
This includes gloves with stimuli on the finger tips~\cite{soviak2016tactile,goncu2011gravvitas,manshad2008multimodal}, computer mice incorporating tactile displays~\cite{rastogi2010issues, jansson2006reading}, force feedback devices~\cite{sjostrom2003phantom, tornil2004use} and friction overlays for touch screens~\cite{xu2011tactile}.
However, these solutions require the use of specialised hardware, which is often expensive and difficult to obtain. 

The vibratory actuators built into consumer mobile devices have been studied as a low-cost alternative to traditional tactile diagrams~\cite{10.1145/3403933, klatzky2014touch}.
One disadvantage of this approach is the associated limited bandwidth, since these vibratory actuators are limited to one contact point.
It has been shown that multiple contact points are necessary for effective path tracing, as the user is then able to judge the curvature of the path and thereby judge the direction to move in order to continue tracing~\cite{rosenbaum2006haptic}.
By using distinct vibratory patterns and sound to denote different components of the diagram, the limited bandwidth of commercial vibratory actuators can be augmented.
However, this again requires the diagram to be enriched with semantic information, and is therefore not useful for existing technical documents.

\subsubsection{Auditory feedback}
Sensory substitution via the sense of hearing has been studied by Meijer~\cite{meijer1992experimental}, who described an algorithm called the vOICe (the capital letters meaning ``Oh, I see'').
The vOICe translates an image to sound by scanning it from left to right while generating a series of tone chords that correspond to the pixel columns of the image. 
Each chord consists of several sinusoids perceived as tones, where each tone indicates a bright pixel in the original image.
The frequency and therefore the pitch of a tone is determined by the vertical position of the pixel in the source image, with higher pixels producing higher-pitched tones. 
Synthesised chords are also placed in the stereo field to indicate horizontal position~\cite{meijer2002seeing}.
Therefore, the leftmost column of the image would be placed to the left of the stereo field while the rightmost column would be placed to the right. 
Alternative image to sound mappings have been described by Capelle \emph{et al.}~\cite{psva} and by Abboud \emph{et al.}~\cite{abboud2014eyemusic}.

Although sensory substitution approaches to graphical accessibility have most often been applied to diagrams where the shape itself needs to be conveyed, it has also been studied as an alternative to the linear approache usually used to represent semantic spatial information like that present in mathematical equations. 
DotsPlus is a system for representing mathematical content as a tactile diagram interlaced with braille symbols~\cite{gardner1995dotsplus}. 
DotsPlus is intended to be produced directly from visual mathematics on a graphical braille embosser. %
Mathematical symbols are replaced by braille counterparts, while visual elements are directly translated to a tactile form.
In this way, most of the spatial layout and graphical indicators of the original visual representation are preserved.

In our own previous work, we extended the vOICe algorithm to allow the interactive exploration of simple diagrams using gestures and a touch screen~\cite{kruger2020interactive}.
This method was shown to allow blind and visually-impaired readers to access
diagrams using a process of interactive interrogation that used the vOICe to ``sonify'' local portions of the diagram selected by finger gestures.
Importantly, this approach allowed the readers to determine and understand the spatial relationships in the diagram.
In this work, we will apply aspects of this technique to the particular case of mathematical equations.
For brevity, we will continue to refer to the rendition of a certain portion an image by the vOICe as its ``sonification''.

\subsection{Accessible games}
The virtual worlds used in accessible games are another setting which requires the representation of complex spatial information in a non-visual way. 
We will use some techniques employed by accessible games as an alternative method for representing the structured spatial information in mathematical equations~\cite{balan2015navigational}.

Games that are accessible to blind and visually impaired users can be broadly categorised into two types, based on the medium of output.
First, there are textual games, the output of which can be accessed by a screen reader~\cite{montfort2012interactive}. 
These games are also referred to as text adventures or interactive fiction. 
Second, there are audio games, which use audio as the primary output modality~\cite{friberg2004audio}.

Most textual virtual worlds employ similar mechanisms to convey spatial layout~\cite{montfort2005twisty}. 
Worlds are divided into locations (known as ``rooms''), which can conceptually represent any amount of physical space. 
Relations between rooms are described by means of exits, which are usually indicated by compass directions, such as ``north'' and ``east'', or ``up'' and ``right''. 
The player is able to explore the environment by issuing textual commands to affect or query the state of the world model. 
The work by Murillo-Morales and Miesenberger~\cite{murillo2020audial} on accessible diagram exploration via natural queries bears some similarity to the interaction modality used in textual games.

Audio games usually provide a number of tools that players can use to explore the environment~\cite{trewin2008powerup}. 
These include simulated sonar (which produces a beep for every object located within a defined range of the listener), footstep sounds that echo in empty spaces, and speech synthesis that describes aspects of the world not easily conveyed by sounds. 
One common functionality is a command which, when triggered, describes an object in a specific direction relative to the player. 
Players may, for instance, request the name of the object in front of them, or the object to their left or right.
Our interest in the methods employed by text games is to explore the use of a direct translation method for reading mathematical equations.
We believe that direct translation facilitated by sensory substitution, combined with the relational exploration conventions of accessible games, may provide some attractive advantages when used to supplement linear representations of mathematical equations. 
First and foremost, direct translation offers a method for accessing content that would otherwise not be accessible to a blind reader, because a linear representation of the content is not available for the overwhelming majority of currently published scientific and technical material.
Second, this approach might facilitate communication between blind and sighted colleagues because it provides the blind reader access to the same spatial representation available to their sighted counterparts.
Finally, it would allow blind authors to verify the visual representation of their work after typesetting it with a linear markup language like braille mathematics or LaTeX, and also serve as an accessible way of verifying the correctness of linear representations automatically produced by software like InftyReader.

\section{Proposed Approach}
Despite many initiatives and standards that aim to improve the accessibility of scientific and technical documents, the reality unfortunately remains that most such publications are only available in PDF format without any accessible markup. 
These documents consist of plain text interleaved with vector drawing instructions (to render equations and vector drawings) and embedded rasterised images (to render bitmap images). 
While such documents can be read using a screen reader, only the plain text portions are accessible. 
Reading the text embedded in the equations of these documents is of very limited use to a blind reader, since the text on its own cannot be used to interpret the equations without graphical information to provide the necessary context.
Even when PDF documents are tagged, they usually do not include sufficient semantic information. 
The PDF document of a technical paper may, for instance, include tags denoting tables and headings, but no tags denoting the semantic information of equations. 
Hence the current reality is that, although screen readers provide access to the textual material present in technical PDF documents, the information contained in equations remains overwhelmingly inaccessible.

To address these limitations, we have developed a procedure to convey the graphical information contained in a mathematical equation by way of audio-visual sensory substitution, while continuing to allow access to symbols and other textual elements via synthesised speech or braille.
Our approach is implemented as a non-visual content browser for graphical information that is interlaced with text, such as a mathematical equation.
This browser draws on the techniques used to navigate and map virtual worlds in text adventure games in order to represent the spatial relationships between elements of the equation.
However it also allows graphical information that cannot be identified using synthesised speech to be explored using an adaptation of the vOICe sensory substitution algorithm first described by Meijer~\cite{meijer1992experimental} and subsequently adapted in our own work to allow the focussed exploration of graphical content in electronic documents~\cite{kruger2020interactive}.  
For the current study, we focus specifically on access to mathematical equations that are graphically represented in inaccessible PDF documents. However, our proposed approach is in principle also applicable to other two-dimensional representations of information, such as charts or infographics.

Our browser provides blind readers with two interface modes, both of which allow the two-dimensional structure of an equation to be explored.
The elements of the equation are considered to be a kind of map of the virtual world being explored.
The reader is always positioned at one of the elements of the equation, referred to as the \textbf{focus}.
The focus is always a textual item that can, for example, be spoken as synthesised speech.
Because the geometric arrangement of the equation is preserved, the reader is able to discover how items are arranged relative to each other.

Non-textual components of the equation, such as fraction lines, square roots and most brackets, are considered to be graphical information and their geometric shape is conveyed using the vOICe algorithm.
Since these non-textual graphical parts of an equation are represented in PDF documents as graphical instructions for which the meaning can not be easily inferred and therefore cannot be rendered as synthesised speech, they can never be in focus and are referred to as ``not navigable''.
Instead, they must be explored from the vantage of the surrounding navigable elements.

\emph{Text mode} enables exploration using textual commands in a style similar to that used to explore the virtual worlds of text adventure games. 
There are commands to examine the focus, and also to move to elements adjacent to the focus.
Output is provided in the form of verbose textual descriptions, containing information about the elements of the equation currently in focus and their spatial placement. 

\emph{Graphical mode} allows the user to explore the equation by using keys on the keyboard, as well as gestures on a touch screen, if one is available.
This interface therefore does not rely on explicit textual commands to be typed in order to move around the equation. 
Textual content that is encountered is announced by the screen reader, while graphical content is rendered as audio interactively by the adapted vOICe algorithm. 

Our approach requires two sources of data: a rasterised image of the document as it would appear on-screen, and the set of constituent textual elements with their locations and bounding rectangles.
The textual elements and locations are obtained automatically by parsing the PDF document using a PDF parser like Poppler~\cite{poppler}.
To facilitate navigation, a document model of the equation under consideration is constructed using all textual elements that are identified by the PDF parser.

\subsection{Document Object Model}
The document object model (DOM) consists of a graph constructed using the navigable elements that can be identified from the source equation.
Non-navigable elements are not contained within the DOM, but are extracted from the rasterised rendering of the equation before sonification (described in Section~\ref{sec:sonification}).
For each navigable element, a node is created containing as data the text of the element as well as its location and bounding rectangle.
Thereafter, edges are added between neighbouring nodes. 
Each node may have up to 12 connected edges, representing four primary directions (left, up, right and down), each with three secondary variants such as up-left, up-centre, and up-right.

To compute the edges, a few simple rules are followed:

\begin{itemize}
  \item A node may be connected to another node by only one edge of a certain kind.
  \item The possible neighbours of a node are prioritised by shortest distance.
  \item An edge between two nodes is only inserted if it does not cross any other nodes.
\end{itemize}

The pseudocode of the algorithm used to construct the edges of a DOM is given in Figure~\ref{fig:domcode}.

\begin{figure}
  \begin{lstlisting}
  for each textual element 1 in input_list {
    for each textual element 2 in input_list (skipping textual element 1) {
      if textual elements 1 and 2 are not within line-of-sight {
        continue;
      }
      let dir be the calculated direction from textual element 1 to 2 (up, up right, etc);
      if textual element 1 has no connection of direction dir, or if element 2 is closer {
        set connection with direction dir from textual element 1 to textual element 2;
      }
    }
  }
\end{lstlisting}

\caption{Pseudocode of the algorithm used to construct the connections of the document object model (DOM).}
\label{fig:domcode}
\end{figure}

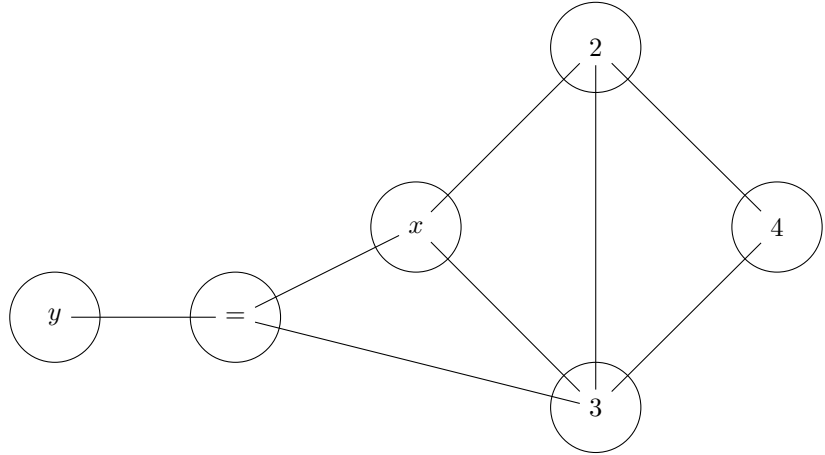
\begin{figure}
  \centering
\begin{tikzpicture}  
  \draw (0,0) circle [radius=0.6cm] node (y) {$y$};
  \draw (2.4,0) circle [radius=0.6cm] node (eq) {$=$};
  \draw (4.8,1.2) circle [radius=0.6cm] node (x) {$x$};
  \draw (7.19,-1.2) circle [radius=0.6cm] node (three) {$3$};
  \draw (7.19,3.59) circle [radius=0.6cm] node (two) {$2$};
    \draw (9.6,1.2) circle [radius=0.6cm] node (four) {$4$};
  \draw (y) -- (eq);
  \draw (eq) -- (x);
  \draw (eq) -- (three);
  \draw (x) -- (two);
  \draw (two) -- (four);
  \draw (x) -- (three);
  \draw (two) -- (three);
    \draw (four) -- (three);
  \end{tikzpicture}
\caption{Document Object Model for $y = \frac{x^2+4}{3}$.}
\label{figdom}
\end{figure}

As an example, consider the equation $y = \frac{x^2+4}{3}$, for which the DOM is shown in Figure~\ref{figdom}.
The DOM contains only textual elements, purely graphical elements are sonified directly from the rasterised rendition of the equation. 
Our algorithm for producing a DOM is deterministic, therefore any given equation will always result in the same DOM.

Note that non-navigable elements like fraction lines and the extent of square roots are not contained within the DOM, and are therefore not taken into account when edges are inserted. These elements are extracted from the rasterised rendering of the equation at the time of sonification.

\subsection{Sonification and Navigation} \label{sec:sonification}
Both text mode and graphical mode allow the user to move among the navigable  elements of an equation.
The focus, which is initially the top left navigable element of the equation, is shared between both interface modes.
Since graphical elements of the equation are not navigable and can therefore never be the focus, our approach is to allow users to interpret such graphical elements by means of audio-visual sensory substitution, using the surrounding navigable textual elements as context.
From the current focus, sonification of graphical content can be requested in the four primary directions.
For instance, if the current focus is the symbol $x$, the reader can request any graphical content directly above this $x$ to be sonified.

Sonification of non-textual graphical content is accomplished by constructing a bounding box located in the desired direction relative to the focus element. 
The bounding box extends from the edge of the focus in the requested direction, up to the edge of any element within line of sight in the same direction, or the edge of the screen. 
For sonification above or below the focus %
the left and right edges of the bounding box are aligned with the edges of the focus. %
However, for sonification to the left or right of the focus, the bounding box extends vertically to include lines with non-zero pixels.
Therefore, sonifying graphical content to the left or right of the focus will usually play the entire graphical element, while sonifying above or below will only play the part directly in line with the focus. 
This allows the user to hear symbols like brackets in their entirety, but elements like fraction lines and the extents of square roots in discrete segments relative to navigable elements. 
The reason for this is two-fold:
\begin{enumerate}
\item Initial testing suggested that a bracket is difficult to interpret when this symbol is only partly sonified, while fraction lines and the extents of square roots are not. 
This is because brackets include much of their recognisable shape in the vertical dimension, which is rendered concurrently by the vOICe, while fraction lines contain information only in the horizontal direction which is rendered as time.
\item For brackets it is sufficient for users to know whether they are present next to a navigable element or not, but for fraction lines and the extents of square roots it may be important to know  exactly which part is above or below a navigable element. The latter is particularly important when more than one graphical element is above or below a sequence of navigable elements. 
For example, consider the equation $y=\frac{3}{\sqrt{x}+2}$. The two elements  $x$ and $2$ are both below the fraction line, but $x$ is also below the extent of the square root. 
In order to correctly interpret the equation, the reader must be able to tell that $x$ has two lines above it, but $2$ has only one.
\end{enumerate}

\subsection{Text Mode}
Text mode browsing views the constituent elements of an equation like locations in a text-based adventure game.
The equation can be explored by issuing text commands to, for example, inspect the current focus, to move left, right, up or down from the current focus, or to describe any graphical elements that surround the focus.
Each command results in informative textual output that describes a particular aspect of the equation. 
This output is conveyed to the reader as synthesised speech or braille, facilitated by a screen reader.
The structure and spirit of these textual commands are deliberately intended to loosely follow the conventions established by textual virtual worlds.
To speed navigation, the responses to many commands include hyperlinks, which can be activated to trigger follow-on commands relevant to the context.

A typical command consists of a keyword describing the action to be performed, followed by an optional argument, for example the name of an element or a direction. 
If a command operates on an element, but no element is specified, the command is assumed to refer to the focus. 
For example, the ``play'' command, when followed by the direction ``right'' will sonify the element immediately to the right of the focus. 
However, when no element is specified, the element currently in focus is sonified.

Movement is facilitated by commands describing the direction, such as ``left'' or ``up''. 
A secondary direction may be specified to refine the movement. 
For example, the ``right up'' command will move the focus diagonally right and up from the current focus.
The ``look'' command allows users to obtain a description of the current focus, or a specific element if provided as an argument. 
This command is also automatically issued when the navigation begins, to provide users with an immediate overview of the initial focus.

The following information is included in an element's description:
\begin{itemize}
\item The textual content of the navigable element, for example, a mathematical symbol (like "$\Omega$") or the text string (like "exp"). %
\item The location of the element within the equation,  as normalised horizontal and vertical Cartesian coordinates between the values of 0 and 100.
\item A list of adjacent elements, with their respective relative directions. These are rendered as links which, when activated, move the focus to that element. An example of such output might be ``From this element, the following elements can be reached: Up: "$x$", Right, "$2$".'' %
\item A list describing whether non-navigable graphical content can be sonified in each possible direction. An example of such output might be, ``There is additional graphical content left, up, and right''. The items in this list are hyperlinks which sonify the relevant graphical content when activated.
\item A link which can be activated to sonify the navigable element being described %
\end{itemize}

In addition, when the ``look'' command is issued without arguments, the output also contains a list of items that are vertically aligned with the  focus.
These horizontally-sequential elements are often terms of an expression.
Users are also notified when the current line changes by a notification of the number of elements in the new line, for example after a movement command.

\subsection{Graphical Mode}
As an alternative to the text-driven exploration described in the previous section, graphical mode allows the user to explore the elements of an equation using the keyboard cursor keys to trigger movement. 
As the user navigates to a new element, it is announced by the screen reader after which the shape of the element is sonified using the vOICe algorithm.
The reader is also informed when the vertical position of the focus changes as a result of a horizontal movement, or when the horizontal position changes as a result of a vertical movement. 
For example, when the reader moves to the right, and the new focus is higher than it was previously, as it might be when moving into the numerator of a fraction, this is announced by the word ``raised''. 
By learning to interpret these announcements, the reader can identify the spatial positions of the elements of an equation, such as exponents and fractions.
When the user navigates to an adjacent element, and in the process passes over non-navigable graphical information, that graphical
information is sonified before that of the newly focused element. 

As described above, the document model allows each node to have up to 12 edges, denoting the four primary directions each with three secondary variants. 
However, since there are only four cursor keys on a standard keyboard, the same fine-grained navigation available in text mode is not possible. 
Instead, horizontal movement prioritises nodes in the upward direction, while vertical movement prioritises nodes in the leftward direction. 
This decision was based on initial user feedback which suggested a preference for reading the numerator of a fraction before the denominator when navigating horizontally, as well as a preference for first reading the left-most element in a term when navigating vertically.  
Although the navigation options in graphical mode may appear more limited than those available in text mode, and as a result sometimes require more steps to arrive at an intended destination, this is mitigated by the much greater speed at which cursor-key driven navigation is possible.

When a touch screen is available, graphical mode also allows the user to explore the content using gestures. 
Each keyboard command has an equivalent touch gesture, but a touch screen has the benefit of making a more spontaneous exploration of the content possible.
As the user moves one finger across the screen, the entire column of the image situated beneath the point of contact is sonified as a tone chord using the vOICe algorithm.
In addition, the portion of the column immediately under the fingertip is sonified with greater intensity.
This is intended to allow targeted and interactive graphical exploration of the equation, while still providing context about the geometrically surrounding area.
Finally, a two-finger gesture allows the line segment between the fingertips to acts as a scanner, with the pixels along this line sonified as a tone chord. 
By moving the fingertips closer together or further apart, or by rotating them around each other, the sonification of shapes with any orientation and size can be achieved.
This interactive and localised image exploration is not possible with the classical implementation of the vOICe algorithm.

\subsection{Integrated Browsing}
When browsing, the user is initially presented with text mode. 
This mode appears similar to a typical computer command line interface, with output displayed above a field for input.
The reader may switch to graphical mode at any time.
The command line interface is then replaced with a canvas area in which the graphical form of the equation is rendered, and through which the reader can navigate using the cursor keys. 
Sonification is provided immediately as the focus changes, while textual notifications regarding the focused element are displayed in a status area and presented by the screen reader.

\section{Experimental Evaluation}
Our proposed method of interactive equation exploration was evaluated by 25 test subjects, 11 of whom were blind and 14 of whom were sighted.
Before evaluation, subjects were provided with a training session in which they could familiarise themselves with the method of exploration and the text and keyboard commands.
After training had been completed, each subject completed two evaluation phases.
In the first, candidates were asked to consecutively identify a number of different equations using only text mode.
In the second phase, candidates were permitted to use both text and graphical modes, and to switch freely between them. 
The exploration algorithm, as well as the training and testing procedures, were implemented using web technologies and were therefore accessible over the internet.
This allowed remote evaluation, which also enabled us to comply with Covid19 safety protocols.

\subsection{Equations under Review}
\label{interpret}

Given that our approach provides access to mathematical content by the selective and structured sonification of graphical (non-textual) structures, we identified a number of mathematical conventions which are difficult to read without access to this information. 
These mathematical structures cannot, for example, be correctly interpreted from the plain-text extracted from an untagged PDF document, as would for example be presented by a screen reader.

The following five print mathematical conventions all depend on graphical structure for their correct interpretation, and are not identifiable when rendered as plain text. While these five conventions have been identified for the purposes of the evaluation we will present, the list is not intedned to be exhaustive.

\begin{itemize}
  \item Exponents. 
  After conversion to plain text, exponents are at best presented over two lines, with the exponent in the first and the base in the second. 
  Fractions are presented in a similar way, resulting in immediate ambiguity.
  \item Fractions. 
  The fraction line is lost during the conversion to plain text. 
  The numerator and denominator are usually presented on separate lines. 
  This is similar to the presentation of exponents, as described above.
  \item Square roots. 
  The radicand is indicated by a graphical line extending over the argument, which is lost during the conversion to plain text.
  Hence it is not possible to determine which terms fall under the root from a plain text representation. 
  Depending on the PDF reader, the square root symbol itself may also be presented on a different line from the radicand.
  \item Large brackets. 
  Depending on the source of the PDF, large brackets are often rendered using fonts with custom glyphs, for which the TTS system cannot identify a corresponding textual description. 
  In print mathematics, the size of a bracket is also often used to convey the relationships between elements of the equation to the reader. 
  Therefore, an accessible two-dimensional view requires the vertical extent of the brackets to be rendered.
  \item Matrices. 
  After conversion to plain text, matrices are presented over several lines, which appear similar to the plain text representation of fractions and exponents. 
  For reasons similar to those described above, the large square brackets delimiting a matrix are also usually absent from the plain text representation.
\end{itemize}

On the basis of these five types of mathematical content,  we developed two sets of six equations with which to evaluate our approach.
The first set was
used to evaluate text mode interaction only, while the second was
used to evaluate both text and graphical modes of interaction.
In both cases, an attempt was made to keep the evaluation short, so as to minimise user fatigue.
We also endeavoured to choose equations that are not identifiable from a plain text representation. 
To achieve this, we chose equations which contain at least two of the graphical conventions in the list above.
The following two tables list the equations  we used for the two stages of our evaluation.

\begin{table}[h]
\begin{center}
\label{tab:equations_stage1}
  \begin{tabular}{|l|l|l|l|l|l|}
    \hline
  Equation & Exponents & Fractions & Roots & Brackets & Matrices \\
\hline
$y=\frac{x}{2}+x^2$ & 1 & 1 & 0 & 0 & 0\\
$y=\sqrt{x}+x^{2}$ & 1 & 0 & 1 & 0 & 0\\
$y=\frac{2}{\sqrt{x}}$ & 0 & 1 & 1 & 0 & 0\\
$y=\frac{\sqrt{x} +x^4}{x -2}$ & 1 & 1 & 1 & 0 & 0\\
$y = \left(\frac{x}{2}+2\right)^2$ & 1 & 1 & 0 & 2 & 0\\ 
&&&&&\\
$
  \begin{bmatrix}
    2 & 4 \\
    2 & 4 \\
  \end{bmatrix}
  \times
  \begin{bmatrix}
    1 \\
    2 \\
    \end{bmatrix}
  $ & 0 & 0 & 0 & 4 & 2\\
  &&&&&\\
\hline
\end{tabular}
\caption{Equations for Stage 1 - Text Mode.}
\end{center}
\end{table}

\begin{table}[h]
\begin{center}
\label{tab:equations_stage2}
  \begin{tabular}{|l|l|l|l|l|l|}
    \hline
  Equation & Exponents & Fractions & Roots & Brackets & Matrices \\
\hline
$y=x^2+\frac{2}{x}$ & 1 & 1 & 0 & 0 & 0\\
$y=x^3+\sqrt{x}$ & 1 & 0 & 1 & 0 & 0 \\
$y=\frac{\sqrt{x}}{2}$ & 0 & 1 & 1 & 0 & 0 \\
$y=\left(\frac{x\sqrt{x}}{x +2}\right)^5$ & 1 & 1 & 1 & 0 & 0 \\
$y = \left(1+\frac{2}{x}\right)^5$ & 1 & 1 & 0 & 2 & 0 \\
&&&&&\\
$
  \begin{bmatrix}
    1 & 2 & 3 \\
    2 & 3 & 4\\
  \end{bmatrix}
  \times
  \begin{bmatrix}
    1 & 2\\
    2 & 3\\
    3 & 4 \\
    \end{bmatrix}
  $ & 0 & 0 & 0 & 4 & 2 \\
&&&&&\\
\hline
\end{tabular}
\caption{Equations for Stage 2 - Text Mode \& Graphical Mode.}
\end{center}
\end{table}

The equations in these two tables were designed to illustrate the current inaccessibility of mathematical content in untagged PDF documents. 
For example, consider Equation~1 from Stage~1:

$$y=\frac{x}{2}+x^2$$

A current mainstream PDF reader extracts the following textual representation when viewing the equation:

\begin{tcolorbox}
\begin{verbatim}
x
2
y =+ x
2
\end{verbatim}
\end{tcolorbox}

We see that the equation is represented as text displayed over several lines as a result of the vertical placement of terms.
It is, however, not possible to determine from this representation which elements are fractions and which are exponents.
It is also not possible to determine the extent of the fraction line, and therefore which terms are part of the numerator and which are part of the denominator. 
For example, the following are all legitimate interpretations of the above textual representation.

\begin{itemize}
\item[] $y = \frac{2}{x}+\frac{2}{x}$
\item[] $y = \frac{2}{x}+x^2$
\item[] $y = \frac{2}{x+x^2}$
\item[] $y = x+\frac{2}{x^2}$
\end{itemize}

As a second example, let us consider Equation~5 from Stage~2.

$$y=\left(\frac{x\sqrt{x}}{x +2}\right)^5$$

The plain text representation of this equation is as follows:

\begin{tcolorbox}
\begin{verbatim}
√  5
xx
y =
x +2
\end{verbatim}
\end{tcolorbox}

This equation is therefore displayed over four lines, due to the spacing of the fraction and the exponent. 
However, the graphical indicators are absent, and therefore it is impossible for a blind reader to distinguish between the exponent and the fraction. 
If we assume that the equation does not contain brackets, the following interpretations are possible.

\begin{itemize}
\item[] $y = x+\sqrt{\frac{x}{x+2}}^5$
\item[] $y = x^x+\sqrt{\frac{x}{2}}^5$
\item[] $y = x^x+\frac{\sqrt{x}}{2}^5$
\item[] $y = \frac{x\sqrt{x}^5}{x+2}$
\end{itemize}

The above examples are intended to illustrate the inaccessibility of equations in untagged PDF documents.
The high ambiguity of the textual representation that can be extracted from such PDF documents means that this is not a viable means by which blind readers can access mathematical content.
For this reason, and to avoid the unnecessary additional fatigue it would cause test subjects, we did not include this plain text format in our evaluation.

\subsection{Test Candidates}

Evaluation was performed by eleven blind and fourteen sighted human subjects.
Subjects were recruited by word-of-mouth, and in the case of blind subjects also by means of mailing lists for blind STEM practitioners.
The majority of sighted candidates were recruited locally from the Departments of Mathematics and Engineering at the University of Stellenbosch.
However, the majority of blind candidates were recruited internationally.
Candidates were between the ages of 18 and 70 and included students (undergraduate and postgraduate) as well as candidates who were not students but had an undergraduate or a postgraduate qualification.
Both blind and sighted groups included candidates with mathematical and scientific backgrounds, but also candidates from fields outside STEM such as law and journalism.
Table~\ref{tab:characteristics} lists some of the attributes of the test candidates.

\begin{table}[h]
  \begin{tabular}{|l|c|c|}
    \hline
    & Blind & Sighted \\
    \hline
    Undergraduate student & 1 & 1 \\
    Postgraduate student & 1 & 7 \\
    Undergraduate qualification & 4 & 5 \\
    Postgraduate qualification & 8 & 10 \\
    STEM background & 8 & 11 \\
    \hline
  \end{tabular}
  \caption{Some attributes of the test candidates.}
  \label{tab:characteristics}
\end{table}

\subsection{Training}

Since mathematical material remains largely inaccessible to blind readers, it was a challenge to identify suitable blind test candidates. 
Even when willing individuals were identified, it was found that many did not read mathematics on a regular basis and therefore required an introduction to remind them of the key concepts, especially the typical graphical layout of equations that sighted readers are accustomed to. 
Hence this training phase, which was carried out prior to the evaluation, was essential.

An online tutorial was developed to re-familiarise participants with the key attributes of mathematical equations as well as to practice the use of the browsing software. 
The tutorial contained a step-by-step introduction and explanation of the functionality of the proposed algorithm. 
This included the commands available in text mode, the commands and gestures available in graphical mode, and an explanation of the sound and synthesised speech output that could be expected while browsing. 
The tutorial also contained a section that allowed users to listen to sound renderings of example known graphical shapes, including a horizontal line, a diagonal line, and a square root.
This was included to familiarise participants with the method of sonification, with which most were also not acquainted.

As many of the blind and visually impaired candidates had no prior experience with print mathematical notation, descriptions of the graphical symbols that form part of the evaluation were also included in the tutorial.
These symbols included a fraction line, a square root, and both round and square brackets. 
Using the tutorial, users were able to practice exploring increasingly complex equations using both text and graphical modes. 
The tutorial also included a section that allowed users to test their ability to interpret a number of equations.

In addition to the tutorial, interactive training sessions were performed with candidates via online voice communications platforms. 
Users were guided through the initial questions of the tutorial, and had the opportunity to ask questions.

\subsection{Testing Procedure}

The evaluation consisted of two sets of six equations, as described in the previous section. 
The first set was used to evaluate the use of text mode for navigation, while the second was used to evaluate the mixed use of text and graphical modes for interaction, where subjects could switch freely between the two.

Like training, evaluation was designed to function over the internet. 
To achieve this, each equation was delivered as a web page and the browsing software was implemented as an interactive web-based application developed using the Rust and Javascript programming languages. 
The start page of the evaluation contained links to each equation, and also contained instructions for the candidate.
Each link opened the software in a new browser window dedicated to the specific equation, and allowed the subject to explore it. 
After exploration, the browser window could simply be closed to return to the starting page.

Equations were typeset using LaTeX and rendered as a PDF to fill the screen, scaled to a horizontal resolution of 300 pixels, preserving aspect ratio.
Although equations were rendered to a canvas, they were not visibly displayed, thus ensuring a similar testing environment for both sighted and blind participants.
Textual symbols, along with their locations and bounding boxes, were extracted from the PDF documents using the Poppler~\cite{poppler} and PdfMinor~\cite{pdfminer} PDF extraction libraries.
For each equation, a table of the symbols extracted in this way, as well as an image containing the graphical rendering of the equation, was stored. 
Equations were rendered in isolation for the purposes of this study since the aim was to determine the effectiveness of the proposed browsing method.
For an integrated solution, equations can be isolated and extracted from the PDF in which they are embedded, either automatically or based on a selection made by the human reader.

After candidates had completed the training tutorials and declared themselves comfortable using the browsing software, they could proceed to the evaluation stages.
The first phase of the evaluation required the participants to explore and identify the first set of six equations using text mode only.
Since the textual output was displayed on the screen, it could be announced by the screen reader to blind candidates.
Most blind candidates used either the Jaws screen reader developed by Freedom Scientific, or the open source NVDA screen reader developed by NVAccess. However, at least one blind candidate used Linux with the Orca screen reader.
The use of screen readers was not required for sighted subjects. 
In contrast to the blind subjects who were all familiar with the use of screen readers, the sighted subjects were not.
Therefore, requiring the sighted subjects to use screen readers in order to enforce the same test procedure among all participants could negatively affecting results in a way that is not indicative of the effectiveness of the methods under evaluation.

The second phase of the evaluation required the participants to explore and identify the second set of six equations, but allowed them to used both text and graphical exploration modes, and to switch among these two freely.
For graphical mode, textual output consisted of short notifications which were displayed on a status bar readable by sighted participants and announced by the screen reader of blind participants.

To allow later analysis, a record of the actions performed and time spent during evaluation was kept for each test subject.
Candidates were asked to write down, in a separate document and using full English sentences, what they believed each equation to be.
Candidates were requested to be as explicit as possible in describing the structure of the equation, for example noting the start and end of fractions, roots and exponents. 
Candidates who were familiar with LaTeX were invited to provide their answers in that notation.

\section{Scoring}\label{scoring}
The transcriptions of the equations provided by the test subjects differed widely in style.
Some were provided in LaTeX notation, others as ASCII maths, and many as full and descriptive English sentences.
Due to this heterogeneous format, all responses were manually and individually assessed for correctness.

Two figures of merit, both expressed as percentages, were used to score the responses made by the test subjects.
The first, which we will refer to as `completely correct', is based on a binary score for each equation in the test indicating whether it was perfectly correct or not.
However, even when an equation is not transcribed with perfect accuracy, the degree to which it is errorful can vary. 
Therefore a second figure of merit, which we will refer to as the `correctness score', was introduced.
The correctness score is based on the number of symbols and graphical elements in the equation concerned, and is 100\% only when all symbols and graphical elements are both correctly identified and correctly placed.
The correctness score also had the effect of normalising the responses, allowing them to be compared.

As mentioned previously, the equations used in the evaluations were designed to contain at least two problematic elements each. 
Problematic elements are those which cannot be unambiguously identified from a plain text representation derived from an untagged PDF document. 
The correctness score is therefore based on six indicators: the five possible problematic elements and textual symbols. 
For each equation, we calculated the maximum score based on the number of times each of these elements was correctly identified, as well as the number of times it was correctly placed.
For each equation transcribed by a test subject, the correctness score is calculated by subtracting the number of inserted and deleted elements, as well as the number of incorrectly placed elements, from the maximum possible score.

\section{Results and Discussion}
Table~\ref{tab:counttab} reports the number of times each of the six equations in Stages 1 and 2 were transcribed without any error by the 25 test subjects, while Table~\ref{tab:scoretab} reports the corresponding correctness scores.
We see that, across all subjects and both stages of the evaluation, 78.1\% of the responses were perfectly correct (73.5\% and 82.7\% respectively for blind and sighted subjects).
The corresponding correctness score was 95.4\% overall, with 93.3\% and 97.6\% for blind and sighted groups respectively.

\begin{table}
\begin{center}
  \begin{tabular}{|c|r|r|r|r|r|r|}
  	\hline
  	 & \multicolumn{3}{|c}{Stage 1}           & \multicolumn{3}{|c|}{Stage 2}   \\
  	 \cline{2-7}  
  	      Equation   & Blind  & Sighted & Overall & Blind & Sighted & Overall \\ \hline
  	1        & 81.8\%          & 78.6\%           & 80.2\%         & 81.8\%          & 92.9\%           & 87.3\%        \\
  	2        & 90.9\%         & 78.6\%           & 84.7\%         & 100.0\%         & 100.0\%           & 100.0\%        \\
  	3        & 81.8\%          & 78.6\%           & 80.2\%         & 90.9\%         & 100.0\%           & 95.5\%        \\
  	4        & 45.5\%          & 50.0\%            & 47.7\%         & 54.5\%          & 64.3\%            & 59.4\%        \\
  	5        & 63.6\%          & 78.6\%           & 71.1\%         & 72.7\%          & 85.7\%           & 79.2\%        \\
  	6        & 63.6\%          & 92.9\%           & 78.2\%         & 54.5\%          & 92.9\%           & 73.7\%        \\ \hline
  	Average    & 71.2\%         & 76.2\%           & 73.7\%        & 75.8\%         & 89.3\%           & 82.5\%       \\ \hline
  \end{tabular}
  \caption{Number of completely correct responses per equation for both evaluation stages.}
  \label{tab:counttab}
\end{center}  
\end{table}

\vspace{5mm}
\begin{figure}
  \centering
   \begin{tikzpicture}
   \begin{axis}[
	x tick label style={
      /pgf/number format/1000 sep=},
    ylabel=Correct,
	xlabel=Question,
	enlargelimits=0.05,
	legend style={at={(0.5,-0.1)},
	anchor=north,legend columns=-1},
	ybar interval=0.7,
]
\addplot [draw=black,fill=blue!40]
	coordinates {(1, 80.2) (2, 84.7) (3, 80.2) (4, 47.7) (5, 71.1) (6, 78.2) (7,0)};
\addplot [draw=black,fill=black!20]
	coordinates {(1, 87.3) (2, 100.0) (3, 95.5) (4, 59.4) (5, 79.2) (6, 73.7) (7,0)};
	\legend{Stage1, Stage2}
\end{axis}
\end{tikzpicture}
\caption{Number of perfectly correct responses per question for each stage.}
    \label{fig:num_correct}
\end{figure}
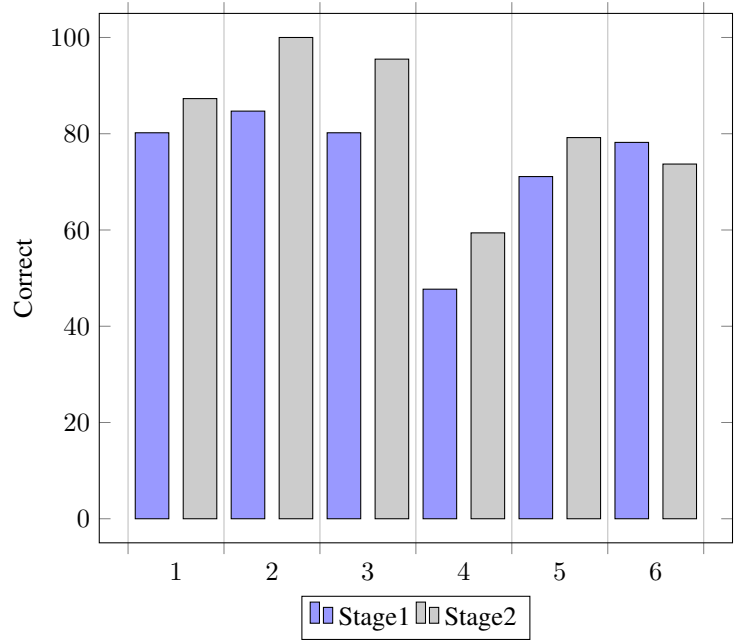
\vspace{5mm}

\begin{table}
\begin{center}
  \begin{tabular}{|c|r|r|r|r|r|r|}
  	\hline
  	 & \multicolumn{3}{|c}{Stage 1} & \multicolumn{3}{|c|}{Stage 2}  \\
  	 \cline{2-7}
  	Equation & Blind   & Sighted & Overall  & Blind    & Sighted  & Overall  \\ \hline
  	1        & 96.0\% & 96.0\% & 96.0\%  & 95.0\%  & 98.4\%  & 96.7\%  \\
  	2        & 98.0\% & 98.8\% & 98.4\%  & 100.0\% & 100.0\% & 100.0\% \\
  	3        & 95.5\% & 97.0\% & 96.2\%  & 99.4\%  & 100.0\% & 99.7\%  \\
  	4        & 88.5\% & 92.6\% & 90.5\%  & 91.9\%  & 95.7\%  & 93.8\%  \\
  	5        & 90.5\% & 95.8\% & 93.1\%  & 93.2\%  & 98.9\%  & 96.0\%  \\
  	6        & 88.8\% & 98.4\% & 93.6\%  & 82.3\%  & 99.4\%  & 90.9\%  \\ \hline
  	Average  & 92.9\% & 96.4\% & 94.6\%  & 93.6\%  & 98.7\%  & 96.2\%  \\ \hline
  \end{tabular}
  \caption{Correctness scores per equation for both evaluation stages}
  \label{tab:scoretab}
\end{center}
\end{table}

Table~\ref{tab:counttab} also shows that, for all except equation four, more than 70\% of the equations were, on average, transcribed perfectly.
Equation four was one of the most difficult equations in both stages, containing all elements identified in Section~\ref{interpret} except matrices.
As expected, the correctness scores in Table~\ref{tab:scoretab} are higher than the corresponding completely correct scores in Table~\ref{tab:counttab}, and are all above 90\%.
This indicates that most elements in an equation were transcribed correctly both in terms of identity and geometric placement.

The tables also show that, even though sighted subjects were not able to see the equation they were exploring, they attained slightly better results on average than blind subjects for both stages of the evaluation.
However, this difference was found to not be statistically significant ($p<0.1$.) 
It must be borne in mind that, although the equation itself was not displayed on the screen, sighted subjects were able to see the textual output generated during browsing. 
Although blind subjects also had access to this information, since the text was synthesised as speech by the screen reader, arguably the ability to read the text allowed sighted candidates to more easily pick out relevant information. 
It should also be highlighted that two blind candidates attained a perfect score on all equations in both stages, which demonstrates the efficacy of our approach even when used entirely non-visually\footnotemark.
\footnotetext{Although not formally analysed, the responses contributed by blind candidates with a STEM background, did not significantly differ from those with other backgounds. For example, of the three candidates without a STEM background, one attained a near-perfect score. We suspect that a general familiarity with technology was ultimately more important.}

Another advantage that sighted subjects arguably have is their familiarity with visual mathematics. 
In performing the evaluation, it was discovered that most blind candidates were unfamiliar with the visual appearance of elements such as square roots, brackets, and matrices, and had to learn these during the training phase, in addition to the use of the browsing software. 
Sighted subjects, on the other hand, were all familiar with the mathematical notation.
Furthermore, sighted subjects were able to form a two-dimensional representation of the equation they were exploring using pen and paper, while blind candidates had to resort to a linear representation like LaTeX or braille to help them.

Tables \ref{tab:counttab} and \ref{tab:scoretab} and Figure~\ref{fig:num_correct} also show that both blind and sighted groups attained higher accuracies in Stage 2, where they could use both text and graphical modes of exploration, than in  Stage 1, in which only text mode could be used.
This difference is statistically significant with $p<0.02$. 
In Stage 1, every question had at least one incorrect response, while for Stage 2, Question 2 was correctly identified by all candidates. 
This may indicate that the graphical mode adds to users' understanding of equations. 
However, all sighted candidates, and ten of the eleven blind candidates also used text mode in Stage 2 for some of the equations, especially for Equations~4 and~6 (the matrix equation).
It should also be remembered that candidates always completed Stage~1 before proceeding to Stage~2, which may also contribute to the higher accuracy attained  in the latter. 

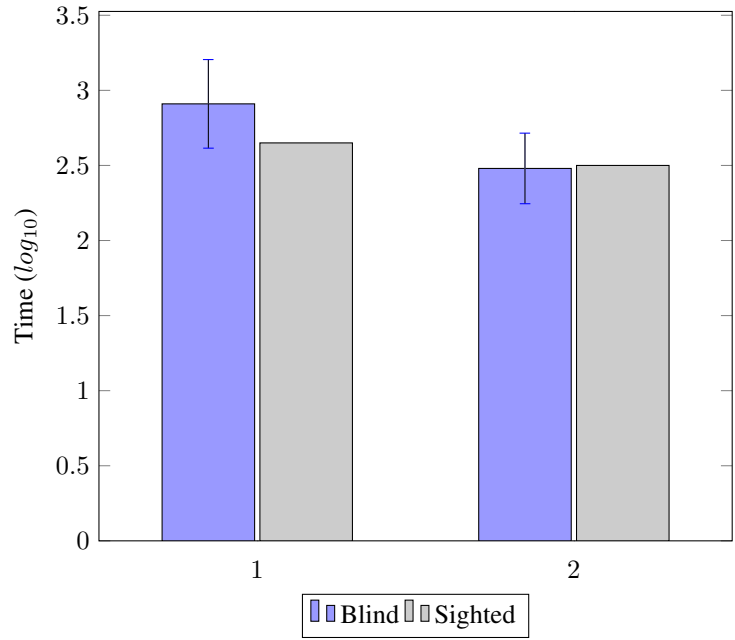
\begin{figure}
    \centering
    \begin{tikzpicture}
    \begin{axis}[
        ybar,
        bar width=35pt,
        xtick distance=1,
        ylabel=Time ($log_{10}$),
        enlarge x limits={abs=0.5},
        legend style={at={(0.5,-0.1)},
	anchor=north,legend columns=-1},
        ymin=0,
        scaled ticks=false,
        xtick style={
            /pgfplots/major tick length=0pt,
        },
    ]
        \addplot+ [draw=black,fill=blue!40,
            error bars/.cd,
                y dir=both,
                y explicit,
        ]
        coordinates {
           (1, 2.91) +- (0,0.295)
           (2, 2.48) +- (0,0.235)
        };
        \addplot+ [draw=black,fill=black!20]  coordinates {
            (1, 2.65)
            (2, 2.50)
        };
        \legend{Blind,Sighted,}
    \end{axis}
\end{tikzpicture}
    \caption{Time taken by blind and sighted users to complete Stages 1 and 2 respectively. Vertical bars denote 95\% confidence intervals.}
    \label{fig:time_diff_stages}
\end{figure}

As can be observed in Figure~\ref{fig:time_diff_stages}, sighted subjects were able to complete Stage~1 more quickly than blind subjects ($p<0.01$), while in Stage 2 the average durations for the two groups were similar.
We believe that the ability to use graphical mode and to switch between this and text mode provided the blind subjects with greater flexibility in how to construct an understanding of the equation structure.
This additional flexibility was less useful to sighted subjects who could make better use of the verbose textual output provided by the browser. %

Sighted candidates also more often used the links in text mode for follow up commands, as apposed to blind candidates, who preferred to type out the commands ($p<0.01$). 
This may also have contributed to the observed time difference between the two groups.

If we examine the results in terms of the particular type of equation that subjects are asked to identify, we see that blind subjects attained lower
scores than sighted subjects for Equation 6 in both phases ($p<0.01$).
Equation 6 included a matrix, and this was incorrectly identified by four blind subjects in Stage 1, and five blind subjects in Stage 2.
However, the same three subjects contributed incorrect responses in both stages.
This suggests that some blind candidates were not familiar with matrices, a hypothesis that is supported by the form of some of the incorrect responses, such as the placement of multiplication sign inside the matrix itself. 
In addition, one candidate omitted the matrix equation entirely in Stage 2. 
The highly two-dimensional structure of a matrix makes it especially difficult to interpret using the linear reading methods commonly available to blind readers. 
In addition, most braille displays can only display one line at a time, and can therefore also not express
the two-dimensional nature of matrices.

Analysis of the answers submitted by blind subjects shows that most errors are due to missing textual elements. 
For instance, for the first equation of Stage~1, two subjects missed the "$2$" in the final exponent. 
For the matrix equations, all but two of the incorrect responses were due to missing textual elements. 
Across all blind subjects, missing symbols accounted for 53\% and 38\% of incorrect answers for Stages~1 and~2 respectively.
It was furthermore observed that subjects often did in fact visit all the elements of the equation during exploration, even though they did not include some of them in their answers. 
This might indicate that candidates missed the elements when writing their answers, rather than when reading them. 
Nevertheless, there were cases where subjects missed elements because they had in fact never visited them during exploration. 
Improving the browser to report whether all textual elements have been visited might go some way in reducing this type of error.

The second most prevalent type of error made by blind subjects was the incorrect placement of brackets. 
Brackets were either not placed in the correct position, or were not included within the submitted answer at all. 
As also seen for symbols, brackets were often omitted even though they had been deliberately sonified by the subject during exploration.
Across all blind subjects, bracket errors accounted for 32\% of incorrect responses for Stage 1, and 31\% for Stage 2.

For sighted subjects, the most prevalent errors were bracket errors, followed by missing symbols.
This finding is interesting, since unlike many symbols that are reflected in the textual output provided to the reader during exploration, brackets are not, and must therefore be rendered as audio.  Hence, for brackets, the sighted subjects do not have the advantage of access to the rendered text.
Across all sighted subjects, bracket errors accounted for 31\% of errors, while missing symbols accounted for 24\%.

The fourth equation was most often incorrectly identified by both groups of candidates in both Stages 1 and 2. 
This equation was arguably one of the most difficult, containing fractions with more than one term in the numerator or denominator with the entire fraction enclosed within brackets.
Most of the incorrect responses to this equation were due to bracket errors.

\section{Informal Feedback}

In addition to the quantitative results gathered during the two testing phases, informal feedback was gathered from the subjects during the training phase as well as after the two testing phases had been completed. 
Overall, the browsing approach was positively received, and most subjects were able to use the interface within an hour of beginning training. 
About half of the blind candidates successfully learnt to use the software by following the online tutorial independently.
For these candidates, individual meetings were scheduled before the testing phases to gather feedback about their initial perceptions, as well as to answer any remaining questions.

Although candidates noted the similarity in functionality provided by the two exploration modes, most blind candidates expressed a clear preference for the graphical mode. 
The graphical mode was perceived to allow faster navigation and an improved grasp of the spatial layout of the equation.
This might be because graphical mode is less verbose and uses quicker (single key) navigational commands. 
One blind candidate noted that the learning curve, and therefore the barrier to entry, was lower for graphical mode.

A number of blind candidates also mentioned that the included links provided in text mode speed up navigation. 
The links were found to be particularly useful when viewing the output in text mode as a history of the steps taken through the equation. 
Specifically, previous output allowed navigational steps to be retraced, and the embedded links could then be used to choose an alternative route
through the equation. 

Although most candidates had no prior experience with the vOICe algorithm, they found the sonification to be intuitive after practice.
One blind candidate reported difficulty in distinguishing between multiple tones when played simultaneously (for example, two parallel horizontal lines), and specifically noted the case of a fraction with a square root in the denominator. 
However, the same candidate was able to identify all equations correctly.

Candidates were asked whether they thought textual elements should automatically be sonified on navigation in exploration mode (the current behaviour), or whether this should be automatic only for non-textual elements (such as fraction lines).
One candidate expressed a clear preference for the latter, with sonification on request. 
However, several other candidates noted that the sonification contributed to their mental image of the equation, and that they preferred automatic sonification.
Because the pitch of the sonification is based on the position of elements on the screen and therefore absolute, removing automatic sonification of textual elements would potentially increase the difficulty in judging weather a fraction line or other element is below or above the current focus. Sonifying the focus along with the graphical element which may be either above or below, allows users to judge the vertical alignment from the relative pitch.

Regardless of their preferred mode of visualisation, eight of the eleven blind candidates reported that the browser provided them with a clearer understanding of the spatial layout used in mathematical equations. 
In particular, one candidate noted that they had not previously had any understanding of the shape of print square roots, and that they were for the first time able to visualise the geometric shape of this symbol after using the browser. 
Another candidate suggested that more semantic information should be added to the textual output, even though they found the spatial information useful. 
It should be borne in mind that our approach was specifically designed to address the accessibility of content with no or insufficient accompanying  semantic information, such as untagged PDF documents. 
Nevertheless, it might prove interesting in future to explore the incorporation of richer semantic textual output, by for example applying our methods to content in MathML or other semantically orientated markup.

\section{Summary and conclusion}

We have introduced a browsing technique specifically aimed at making previously inaccessible mathematical equations, as typically found in 
untagged PDF documents, accessible to blind or low-vision readers.
This browsing technique combines spatial navigational techniques commonly used in text-based adventure games with audio-visual sensory substitution for non-textual elements of the equation.
Two modes of interaction have been implemented and evaluated:
one purely text-based, and the other more directly based on the geometric layout of the equation and that accommodates gestures when a touch screen is available.
During browsing, the reader can switch freely between these two modes of interaction.
For audio-visual sensory substitution, we have adapted and extended the vOICe algorithm, first proposed by Meijer, to allow the interactive exploration of graphical elements in an equation, such as root signs, fraction lines and many types of brackets.
Such elements are represented as graphics in PDF documents, and can therefore not easily be rendered as synthesised speech.

Evaluation of our approach by both blind and sighted test subjects showed that, after a training session in many cases not exceeding one hour, equations that would be inaccessible using current prevalent screen readers can be identified to a high degree of accuracy.
With the exception of one equation in both stages of the evaluation, on average more than 70\% of equations were perfectly identified. %
Indeed, it was possible for subjects from both the blind and the sighted groups to obtain a perfect score.
While sighted candidates expressed a preference for the text-mode interaction, blind candidates made extensive use of the graphical mode, but switched between the two when presented with more complex geometric arrangements such as matrices.
Informal discussions with the blind candidates after the evaluation also revealed that in some cases additional insight into the layout conventions of mathematical equations has been gained.
This indicates that the browsing approach we propose presents a means by which print-disabled readers can, without additional assistance, interactively explore and thereby decipher mathematical content that was, due to its inaccessible format, unfamiliar to them.

The most common type of error made by blind test subjects was the omission of symbols from the equation.
This is an aspect we aim to address in ongoing work, since in some cases the reader can be alerted to the possibility that this type of error is being made.
In addition, we aim to consider different input sources.
Firstly, much older published material is available only in the form of rasterised images embedded in the PDF format, as may be obtained from a document scanner. 
By including optical character recognition (OCR), it may be possible to also make such content accessible using our approach.
Secondly, we would like to consider other non-textual information present in scientific and technical documents, such as graphs and plots.
The overall objective is to develop methods and tools that fill the accessibility gaps in electronic documents that are not currently addressed by screen reading software and that continue to impede access to blind and sight-disabled readers.

\section{Acknowledgements}
The authors would like to thank the South African Council for Scientific and Industrial Research (CSIR) who supported this research. 
We would also like to thank Prof. Martin Kidd from the Centre for Statistical Consultation at the University of Stellenbosch, for assisting with the analysis of the data. Finally, we would like to thank all the test candidates who contributed there time to this study. 

%


%
%
\end{document}